\documentclass[a4paper,11pt]{article}
\pdfoutput=1 
\usepackage{jheppub} 
\usepackage[T1]{fontenc} 

\usepackage[numbers]{natbib}
\usepackage{filecontents}

\def\be{\begin{eqnarray}}
\def\ee{\end{eqnarray}}

\title{\boldmath Universal Statistics of Vortices in a Newborn Holographic Superconductor: Beyond the Kibble-Zurek Mechanism}
\author[a,b,c,d]{Adolfo del Campo,}
\author[b,e]{Fernando Javier G\'omez-Ruiz,}
\author[f]{Zhi-Hong Li}
\author[g]{Chuan-Yin Xia}
\author[h]{Hua-Bi Zeng}
\author[f]{Hai-Qing Zhang}

\affiliation[a]{Department  of  Physics  and  Materials  Science, University  of  Luxembourg,\\L-1511 Luxembourg,  Luxembourg}
\affiliation[b]{Donostia International Physics Center,\\E-20018 San Sebasti{\'a}n, Spain}
\affiliation[c]{IKERBASQUE, Basque Foundation for Science,\\E-48013 Bilbao, Spain}
\affiliation[d]{Department of Physics, University of Massachusetts Boston,\\100 Morrissey Boulevard, Boston, MA 02125}
\affiliation[e]{Departamento de F{\'i}sica, Universidad de los Andes,\\A.A. 4976, Bogot{\'a} D. C., Colombia}
\affiliation[f]{Center for Gravitational Physics, Department of Space Science \& International Research Institute for Multidisciplinary Science, Beihang University,\\Beijing 100191, China}
\affiliation[g]{School of Science, Kunming University of Science and Technology,\\Kunming 650500, China}
\affiliation[h]{Center for Gravitation and Cosmology, College of Physics Science and Technology, Yangzhou University,\\Yangzhou 225009, China}
\emailAdd{adolfo.delcampo@uni.lu}
\emailAdd{fj.gomez34@dipc.org}
\emailAdd{lizhihong@buaa.edu.cn}
\emailAdd{xiachuanyin@163.com}
\emailAdd{zenghbi@163.com}
\emailAdd{hqzhang@buaa.edu.cn}
\abstract{Traversing a continuous phase transition at a finite rate leads to the breakdown of adiabatic dynamics and the formation of topological defects, as predicted by the celebrated Kibble-Zurek mechanism (KZM).  We investigate universal signatures beyond the KZM, by characterizing the distribution of vortices generated in a thermal quench leading to the formation of a holographic superconductor. The full counting statistics of vortices is described by a binomial distribution, in which the mean value is dictated by the KZM and higher-order cumulants share the universal power-law scaling with the quench time. Extreme events associated with large fluctuations no longer exhibit a power-law behavior with the quench time and are characterized by a universal form of the Weibull distribution for different quench rates.}

\begin{document} 
\maketitle
\flushbottom
\section{Introduction}
The dynamics across a continuous phase transition is a paradigmatic scenario of spontaneous symmetry breaking in which adiabaticity inextricably breaks down. In any finite time scale, a quench from the high-symmetry phase to the lower-symmetry phase is governed by critical slowing down and the effective freezing of the system. Facing a degenerate manifold, causally separated regions of the system make disparate choices of the broken symmetry that result in the formation of topological defects~\cite{Kibble76a,Zurek96a}. The characterization of the latter depends on the topology of the vacuum manifold. In the formation of superconductors and superfluids, entailing  $U(1)$ symmetry breaking, vortices with quantized flux appear~\cite{Kibble76b,Zurek96c}. Spontaneously formation of vortices was observed in experiments with neutron-irradiated superfluid Helium~\cite{Ruutu1996,Bauerle1996}.\\
\\
The mean number of topological defects generated in the course of a phase transition is predicted by the KZM to follow a universal power law with the rate at which the phase transition is crossed.
The verification of this prediction has been the subject of a long-time quest~\cite{DZ14}. The validity of the KZM is not only supported by theoretical models and numerical simulations but has been established in a variety of experimental platforms ranging from colloids~\cite{Deutschlander15} to quantum simulators~\cite{Keesling2019,Bando20}.\\
\\
In strongly coupled systems, the validity of KZM cannot be taken for granted. A natural framework to account for strong coupling is provided by holography~\cite{Hartnollbook,Liu_2019}. In this context,  the spontaneous current formation in a superconducting ring~\cite{Sonner:2014tca,ZengXia19,XiaZeng20} is well described by the KZM prediction~\cite{Zurek96a,Das2012}. However, deviations from KZM have been predicted and the power-law scaling of the mean number of topological defects is expected to be modified by logarithmic terms of the quench rate~\cite{Chesler:2014gya}. In the laboratory,  pioneering experiments on the spontaneous vortex formation in the light of the KZM were restricted to the weakly interacting regime, accessible with Bose-Einstein condensates~\cite{Weiler2008,Navon15,Chomaz2015} and ferroelectrics~\cite{Lin2014}.  Remarkably, recent progress has allowed probing the strongly-interacting regime using a unitary Fermi gas~\cite{Ko2019}.  The measured density of defects was found to be compatible with the KZM scaling laws as in the weakly interacting case. Theoretical~\cite{Chesler:2014gya} and experimental~\cite{Ko2019} results on vortex formation at strong coupling are thus in conflict.\\
\\
The predictive power of the KZM is restricted to the average number of topological defects. Spatial correlations between topological defects have been discussed in the framework of the Halperin-Liu-Mazenko theory~\cite{Halperin81,LiuMazenko92}. Fluctuations of the number of topological defects have recently been explored in spin chains~\cite{delCampo:2018hpn,Cui19,Bando20} and one-dimensional $\phi^4$ theory~\cite{Gomez-Ruiz:2019hdw}. These studies have unveiled signatures of universality in the full counting statistics of topological defects that lie beyond the scope of the KZM.\\
\\
In this work, we explore the statistics of vortices in a newborn holographic superconductor in (2+1) dimensions and show that the full counting statistics of vortices is universal. The mean density is shown to follow the KZM power-law prediction. Fluctuations beyond the mean are probed by low-order cumulants of the vortex number distribution, which are found to exhibit a universal power-law scaling with the quench time. The vortex number distribution is well described by a binomial distribution restricted to even outcomes by the topology of the system, making it possible to probe rare events far away from equilibrium.  Large deviations away from the mean vortex number no longer exhibit a power-law behavior and the corresponding extreme value statistics is characterized by a Weibull distribution.
\section{Formation of a newborn holographic superconductor}
We simulate the superconducting transition from a normal metal to a holographic type-II superconductor in two spatial dimensions by implementing a thermal quench in a finite time $\tau_Q$. This results in the spontaneous formation of vortices that are pinned. The system is described making use of the gauge-gravity duality and numerical simulations involving a (3+1) dimensional gravity. In this setup, a thermal quench can be effectively simulated by changing the charge density in the boundary of the black hole, see section~\ref{methods}~\cite{Hartnoll08}. The phase transition is continuous and of second-order. Critical slowing down in the proximity of the critical point leads to vortex formation. To characterize the resulting vortex number distribution, we consider a  homogeneous system, free from external potentials that can alter the KZM scaling~\cite{DRP11}.\\
\\ 
According to the KZM, traversing the phase transition at finite rate leads to the formation of domains of characteristic length scale $\hat{\xi}=\xi_0(\tau_Q/\tau_0)^{\frac{\nu}{1+z\nu}}$, where $\nu$ and $z$ are the correlation-length and dynamical critical exponents of the continuous phase transition, and $\xi_0$ and $\tau_0$ are microscopic constants. Within such domains, the superconductor phase is chosen coherently. According to the geodesic rule~\cite{Kibble76b}, when multiple domains merge at a point, there is a chance that the quantized circulation of the superconductor phase $\phi$ around that point is non-zero and a multiple $2\pi$. This configuration can lead to the formation of a vortex.  Typical values of its vorticity $V=\frac{1}{2\pi}\oint d\gamma \nabla \phi\in \mathbb{Z}$ are  $V=\pm 1$. The number of vortices induced by the thermal quench is thus proportional to $\langle n\rangle\propto A/\hat{\xi}^2$ where $A$ is the area of the superconductor. As a result, the mean vortex number scales with the quench rate following the universal power law $\langle n\rangle\propto(\tau_0/\tau_Q)^{\frac{2\nu}{1+z\nu}}$, which is the key prediction of the KZM. This universal scaling quantifies the intuition that fast quenches result in a high number of vortices, the number of which decreases as the rate of the transition is reduced. As the transition is thermal, the number of vortices is not deterministic but it constitutes a stochastic variable described by a probability distribution.\\
\\
Characterizing the full counting statistics of spontaneously formed vortices across the phase transition is our central goal. As we shall see, fluctuations away from the mean value are universal. Specifically, we show that cumulants of the distribution share with the mean value a universal power-law behavior in the limit of slow quenches, required for scaling theory to hold. In addition, knowledge of the exact vortex number distribution allows us to characterize extreme events associated with large deviations from the mean value.
\section{Methods}\label{methods}
\subsection{Holographic setup}
We work in the black brane background in Eddington-Finkelstein coordinates
\begin{equation}
ds^2 = \frac{L^2}{z^2} (-f(z) dt^2 - 2dtdz + dx^2 + dy^2),
\end{equation}
with $f = 1 - (z/z_h)^3$. The horizon is $z_h$ while $z = 0$ is the boundary in which the field theory lives.  In the probe limit we adopt the Abelian-Higgs Lagrangian density as
\begin{equation}\label{density}
\mathcal{L} = -\frac{1}{4} F_{\mu \nu} F^{\mu \nu} - |D \Psi|^2 - m^2 |\Psi|^2,
\end{equation}
where $D_\mu=\nabla_\mu -iA_\mu$ is the covariant derivative, $A_\mu$ is the $U(1)$ gauge field, $F_{\mu\nu}=\partial_\mu A_\nu-\partial_\nu A_\mu$ is the gauge field strength and $\Psi$ is the scalar field. We take the ansatz as $\Psi = \Psi(t,z,x,y), A_t = A_t(t,z,x,y), A_x = A_x(t,z,x,y), A_y = A_y(t,z,x,y)$ and $A_z = 0$. The equations of motions (EoMs) of these fields can be obtained readily from the Lagrangian density. The explicit forms of the EoMs can be found in the Appendix~\ref{EQMov}. Near the boundary $z\to0$ the expansion of the fields are (we have set $L=1$ and $m^2= -2$), $A_\mu\sim a_\mu+b_\mu z+\mathcal{O}(z^2),~~~\Psi=\Psi_0 z+\Psi_1 z^2+\mathcal{O}(z^3)$.\\ 
\\
 From gauge-gravity duality, the coefficients $a_t, a_i~ (i=x, y)$ and $\Psi_0$  can be interpreted as chemical potential, gauge field velocities, and scalar operator source in the boundary, respectively. Their corresponding conjugate variables are achieved by varying the renormalized on-shell action $S_{\rm ren}$ with respect to these coefficients.  To get a finite $S_{\rm ren}$, some counter terms should be added. According to holographic renormalization~\cite{Skenderis:2002wp}, the counter terms of the scalar field is $S_{\rm ct}=\int d^3x\sqrt{-\gamma}\Psi^*\Psi$, where $\gamma$ is the reduced metric on the $z\to0$ boundary. We have imposed Neumann boundary conditions for gauge fields as $z\to0$ in order to get dynamical gauge fields in the boundary~\cite{witten,silva}. Therefore, a surface term $S_{\rm surf}=\int d^3x\sqrt{-\gamma}n^\mu F_{\mu\nu}A^\nu$, where $n^\mu$ is the normal vector perpendicular to the boundary, should also be added to have a well-defined variation.  Therefore, the expectation value of the order parameter on the boundary,  $\langle O\rangle=\Psi_1$, can be obtained by varying the finite renormalized action $S_{\rm ren}$ with respect to the source term $\Psi_0$. The conservation equation of the charge density and current on the boundary is $\partial_tb_t+\partial_iJ^i=0$, where $b_t=-\rho$ with $\rho$ the charge density, and the current along $i$-direction is $J^i=-b_i-(\partial_ia_t-\partial_ta_i)$.\\ 
\\
To study spontaneously symmetry breaking of the order parameter, we set $\Psi_0=0$ in the $z\to0$ boundary. Besides, the Neumann boundary conditions for gauge fields can be imposed from the above conservation equations. In the spatial $(x, y)$-directions we impose the periodic boundary conditions for all the fields. At the horizon, we set the time component of gauge fields as $A_t(z_h)=0$, and the regular finite boundary conditions for other fields.\\ 
\\
To drive the system out of equilibrium, we quench the charge density $\rho$ on the boundary while fixing the temperature of the black hole, which was commonly implemented in holographic superconductor settings~\cite{Hartnoll08}. The mass dimension of black hole temperature $T$ is one, while the mass dimension of the charge density $\rho$ is two. Thus, $T/\sqrt{\rho}$ is a dimensionless quantity. Therefore, decreasing the temperature is equivalent to increasing the charge density. In order to have a linear quench of temperature across the critical point, one can quench the charge density $\rho$ as $\rho(t)=\rho_c\left(1-t/\tau_Q\right)^{-2}$ with critical charge density as $\rho_c\approx4.06$.
\subsection{Numerical Scheme}
Before quench, we thermalize the system by adding random seeds into the system in the normal state. The random seeds of the fields are added in the bulk by satisfying the statistical distributions $\langle s(t,x_i)\rangle=0$ and $\langle s(t,x_i)s(t',x_j)\rangle=h\delta(t-t')\delta(x_i-x_j)$, with the amplitude  $h\approx10^{-3}$. In principle, $h$ cannot be too large since the seeds serve as fluctuations to thermalize the system.  The system is quenched linearly from $T_i=1.4T_c$ to $T_f=0.8T_c$ with $T_c$ the critical temperature.  We evolve the system by using the fourth-order Runge-Kutta method with a time step $\Delta t=0.02$. In the radial direction $z$, we use the Chebyshev pseudo-spectral method with 20 grids.  Since along $(x, y)$-directions periodic boundary conditions are imposed, we adopt the Fourier decomposition in the $(x, y)$-directions. The size along $(x,y)$ is $50\times50$, and the number of grids are $201\times201$. Filtering methods are implemented following the rule that the uppermost one-third Fourier modes are removed~\cite{Chesler:2013lia}. We count the number of vortices as the average order parameter just arrived at its equilibrium value. Due to the large dimensions of the system (3+1-dim), the time cost in large quench time is considerable. For instance of $\tau_Q=4000$, each trajectory of the simulation will cost more than two hours. Therefore, collecting statistics with 1000 trajectories required about three months of running time. Due to the large consumption of time, we have limited the number of trajectories for each quench time to 1000.
\begin{figure}[t!]
\centering
\includegraphics[width=1.0\linewidth]{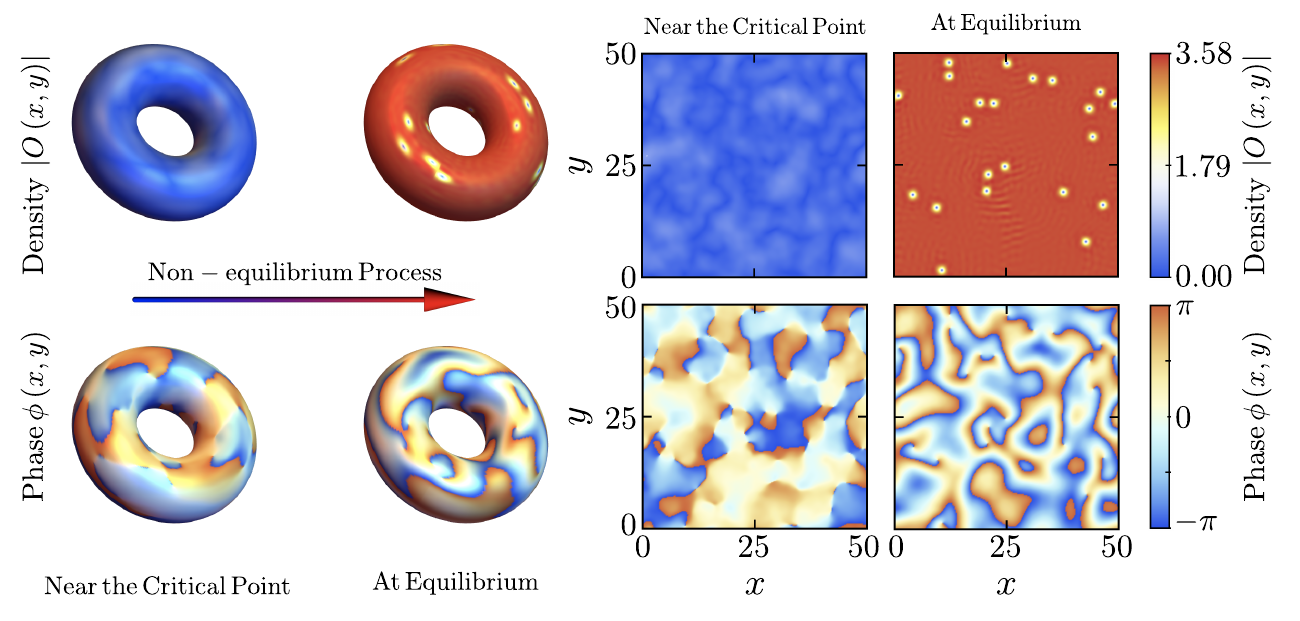}
\caption{(Top) Phase of the superconductor after the thermal quench.  After the non-equilibrium dynamics, the phase in the top panel will evolve into the bottom panel (phase ordering). Now it is hard to see the ``domain structure'' like the bottom panel. There will be a branch cut connecting two defects if the phase geometry is a closed space. (Bottom) Phase of the order parameter near the critical point. The domain structures of the phase space are shown. Among them, there are singular points (topological defects) where vortices will eventually form.}\label{FigPhase}
\end{figure}
\section{Vortex counting statistics}
\begin{figure*}[t!]
\centering
\includegraphics[width=1.0\linewidth]{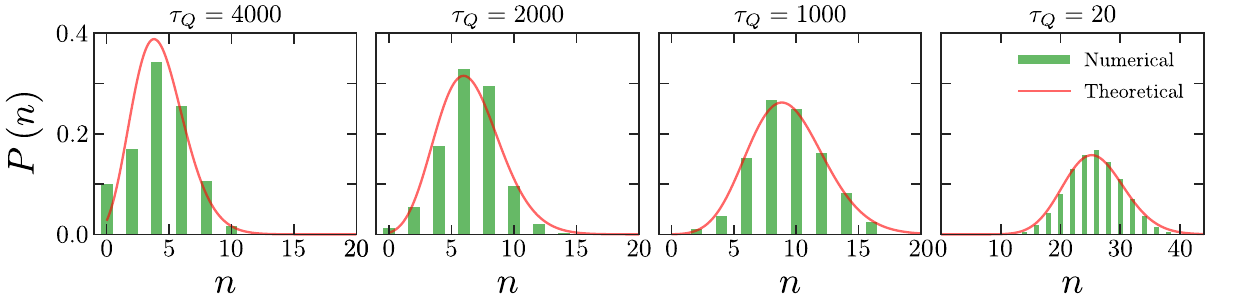}
\caption{Histogram for the vortex-number probability distribution for various quench times. The distribution is asymmetric at the onset of adiabatic dynamics and becomes symmetric for faster thermal quenches that increase the average vortex number. Numerical simulations are well described by the EP distribution in Eq. (\ref{peven1}) with the same average vortex number $\langle n\rangle$.  
The size of the system is $50\times50$ and the number of trajectories for each $\tau_Q$ is 1000.}\label{FigHistpdf}
\end{figure*}

We consider the formation of a newborn superconductor in two spatial dimensions with periodic boundary conditions in the $(x,y)$ spatial directions. The topology of the system is that of a torus $\mathbf{T}^2$ with zero Euler characteristics $\chi(\mathbf{T}^2)=0$. 
As a result of the Poincar\'e-Hopf theorem \cite{Laver10}, the total vorticity of the superconductor equals $\chi$ and vanishes identically.  By contrast, in the case of open boundary conditions, the vortex number would be unrestricted by the topology of the system and could take both even and odd values.
The regime of parameters studied is such that no vortex with vorticity other than $V=\pm 1$ is observed, and the number of vortices and anti-vortices is thus balanced, see Figure \ref{FigPhase}.
We focus on the total number of vortices, regardless of their vorticity. 
By numerically solving the stochastic dynamics in the holographic setting,  an ensemble of realizations is used to collect statistics and build a histogram for different values of the vortex number. For increasingly fast quenches ($\tau_Q=4000, 2000, 1000$ and $20$) the distribution shifts to higher mean values, while simultaneously broadening. By contrast, at the onset of adiabatic dynamics the distribution narrows down and becomes increasingly asymmetric, as the fully adiabatic regime with $n=0$ acquires a significant probability $P(0)$; see Fig.~\ref{FigHistpdf}. 
The vortex number statistics is found to be precisely described by a binomial distribution restricted to even outcomes, which we refer to as an even-binomial (EB) distribution in the following.
The later results from assuming that the probability for vortex formation at the merging point between adjacent domains occurs with probability $p$, while no vortex is formed at such location with probability $(1-p)$. With only two possible outcomes this process can be described as a single Bernoulli trial. The total number of vortices is the result of a number of trials $N\sim A/\hat{\xi}^2$. We next assume that vortex formation at different locations unfolds as a result of uncorrelated events that can be described by $N$ independently and identically distributed (iid) random variables.
The probability to observe a given vortex number $n$ is thus
\begin{eqnarray}
P_{\rm EB}(n)=\frac{1}{A}\binom{N}{n}p^n(1-p)^{N-n},
\label{EBD}
\end{eqnarray}
for any even integer $n\geq 0$, with  $A=\frac{1+(1-2p)^N}{2}$ as normalization constant.
Signatures of universal critical dynamics are encoded in the estimate of the number of Bernoulli trials for vortex formation $N$, while the nature of the vortex number fluctuations is determined by the stochastic model which determines the shape of the distribution.
For small values of $p$,
Le Cam's theorem \cite{LeCam60} guarantees that the statistics becomes  even-Poissonian (EP)
\be\label{peven1}
P_{\rm EP}(n)={\rm sech}(\lambda)\frac{\lambda^n}{n!},
\ee
where $\lambda=Np$ is a parameter and the mean reads $\langle n\rangle =\lambda \tanh(\lambda)$. 
Additional supporting evidence of the excellent  agreement between the $P(n)$  values  extracted from the histogram and those predicted by the distribution (\ref{peven1}) is provided in the Appendix~\ref{BinoEven}  for different values of $n$.
\begin{figure}[b]
\centering
\includegraphics[width=0.8\linewidth]{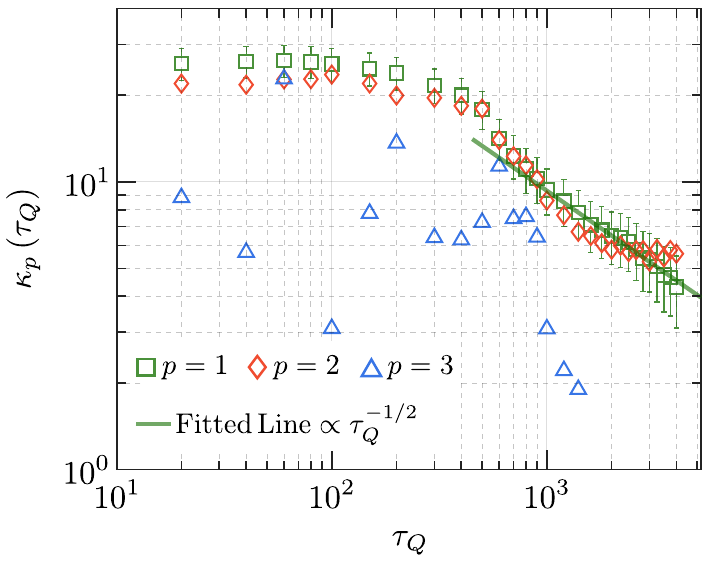}
\caption{Universal scaling of low-order cumulants of the vortex number distribution.  The  solid line is the fitting of the $\kappa_1$, with power $\kappa_1\sim \tau_Q^{-1/2}$. The error bars stand for standard deviations. Within error bars, the variance of the vortex number quantified by the second cumulant exhibits the same power-law scaling and its magnitude is consistent with that of  the mean vortex number, indicating Poissonian statistics. Finite sampling statistics generally prevents establishing the universal scaling of high-order cumulants. \label{FigCumulants}}
\end{figure}

To characterize universal signatures in the full counting statistics, we describe the scaling of the low-order cumulants of the distribution as a function of the quench time. Given the Fourier transform of the distribution $\tilde{P}(w)=\mathbb{E}[e^{iwn}]$, where $w$ is variable conjugated to $n$, cumulants $\kappa_p$ are defined through the expansion $\log \tilde{P}(w)=\sum_{p=1}^\infty\kappa_p(iw)^p/p!$.
The first cumulant equals the average vortex number $\langle n\rangle$ and is predicted by the KZM. The second cumulant equals the variance $\kappa_2={\rm Var}(n)$ while the third one is related to the skewness ${\rm Skew}(n)$ of the distribution through the identity $\kappa_3={\rm Skew}(n)\kappa _{2}^{3/2}$.
In the Poissonian limit for large average vortex numbers, all cumulants approach the first, 
\begin{eqnarray}
\kappa_p\rightarrow \langle  n\rangle,
\end{eqnarray}
thus inheriting the universal power-law scaling as a function of the quench time dictated by KZM.
This prediction is explicitly verified in Fig. \ref{FigCumulants} where the first three cumulants are plotted as a function of the quench time using about 1000 trajectories. This sampling size is limited by the computational cost (see \hyperref[methods]{Methods}).  The first two cumulants show saturation at a plateau for fast quenches, followed by a universal power-law behavior for longer values of the quench time.
The scaling of the first cumulant is dictated by the KZM to follow a power-law $ \langle n\rangle\propto \tau_Q^{-1/2}$ for mean-field critical exponents $\nu=1/2$ and $z=2$ in two spatial dimensions. A fit to the data shows that  $\langle n\rangle\propto \tau_Q^{-0.518\pm 0.0243}$. 
The even Poissonian distribution for a large mean number of vortices predicts a power-law scaling of the vortex number variance $\kappa_2={\rm Var}(n)\propto \tau_Q^{-1/2}$, i.e., equal to the KZM scaling for the mean.
The large fluctuations observed in the third cumulant are expected for the number of trajectories considered and their suppression would require one to increase the number of trajectories by one to two orders of magnitude. 
\section{Large fluctuations}
\begin{figure}[t]
\centering
\includegraphics[width=1\linewidth]{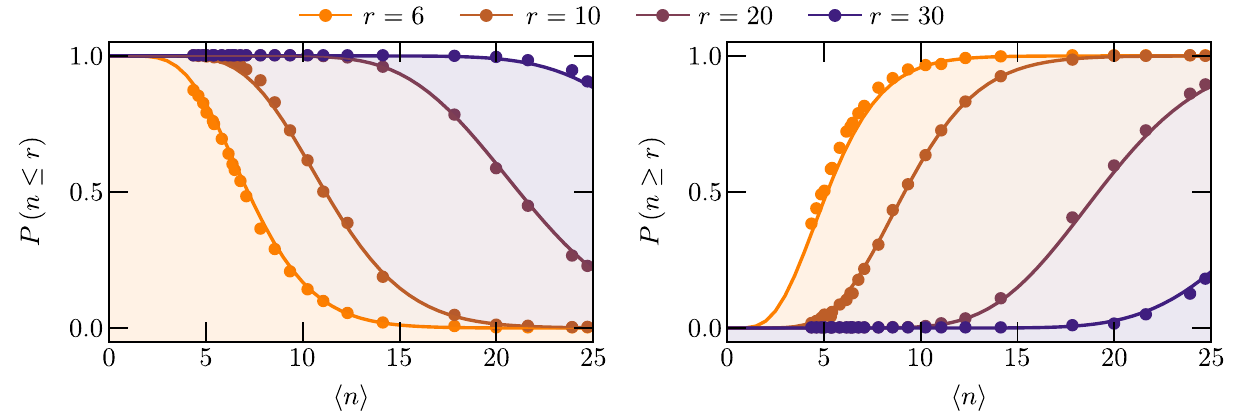}
\caption{Cumulative probability for $P(n\leq r)$ (Left) and $P(n\geq r)$ (Right) in the tails of the vortex number distribution, with respect to the average vortex number $\langle n\rangle$. Numerical dots obtained for different quench times match the theoretical predictions Eq. \eqref{pebksr1} and Eq. \eqref{pebkgr1} (solid lines) derived from an EP distribution. }\label{pkrn}
\end{figure}

In what follows we turn our attention to extreme statistics associated with rare events, that can be estimated efficiently with the available sample size. Fluctuations far away from the mean vortex number can be expected to be sensitive to defect-defect interactions.
Their characterization can be achieved by adding the contribution from the tails of the distribution.
In the even-Poissonian limit, 
the cumulative probability for $P_{ {\rm EP}}(n\leq r)$ and 
$P_{\rm EP}(n\geq r)$ are

\be\label{pebksr1}
P_{\rm EP}(n\leq r)&=&1-\frac{\text{sech}(\lambda ) \lambda ^{2
   \left\lfloor \frac{r}{2}\right\rfloor +2} \,
   _1F_2\left(1;\left\lfloor \frac{r}{2}\right\rfloor
   +\frac{3}{2},\left\lfloor \frac{r}{2}\right\rfloor
   +2;\frac{\lambda ^2}{4}\right)}{\left(2
   \left(\left\lfloor \frac{r}{2}\right\rfloor
   +1\right)\right)!}, \\
P_{\rm EP}(n\geq r)&=& \frac{\text{sech}(\lambda ) \lambda ^r \,
   _1F_2\left(1;\frac{r}{2}+\frac{1}{2},\frac{r}{2}+1
   ;\frac{\lambda ^2}{4}\right)}{r!}. \label{pebkgr1}
\ee

in which $\, _1F_2$ is a hypergeometric function. For the values of $r=6, 10, 20$ and $30$, the cumulative probability is shown in Fig. \ref{pkrn} as a function of the mean number with an excellent agreement between the numerical data and the prediction for the even Poissonian distribution. 
\begin{figure}[h]
\centering
\includegraphics[width=1\linewidth]{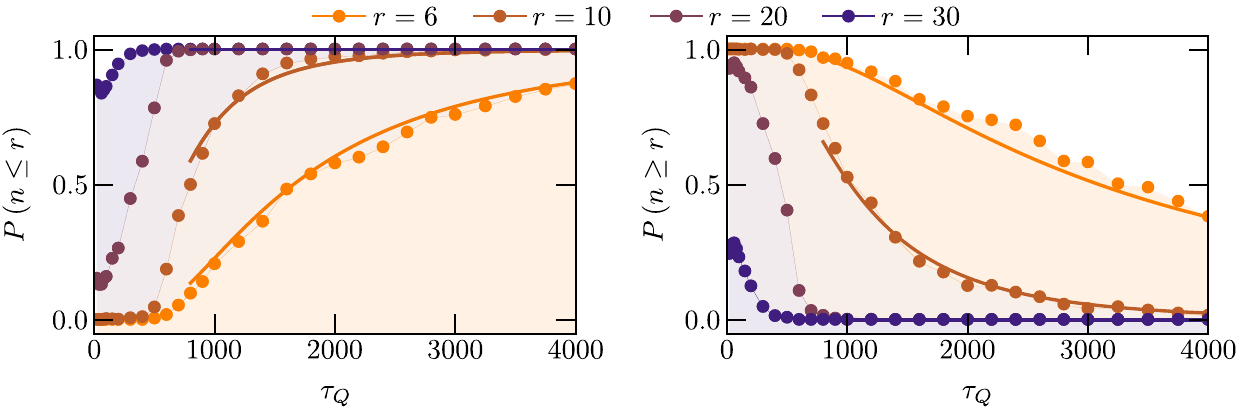}
\caption{Universal scaling in the tails of the vortex number distribution. The cumulative probabilities  $P(n\leq r)$ (Left) and $P(n\geq r)$ (Right) are shown as functions of the quench rate $\tau_Q$. The analytical results (solid lines) for the EP distribution are shown in the  range of quench times from $\tau_Q=800$ to $\tau_Q=4000$, where the numerical simulations exhibit Kibble-Zurek scaling. }\label{pkrtauq}
\end{figure}
As $\langle n\rangle$ is predicted by the KZM, it can be fitted to power the scaling of  $\langle n\rangle \approx 333.3\times \tau_Q^{-0.518}$
in the range of quench times $\tau_Q\in[1000,4000]$ (see Fig. \ref{FigCumulants}), together with  Eqs. \eqref{pebksr1} and \eqref{pebkgr1}, to quantify the quench time dependence of rare events, shown  in Fig. \ref{pkrtauq}. 

\begin{figure}[t]
\centering
\includegraphics[width=1.0\linewidth]{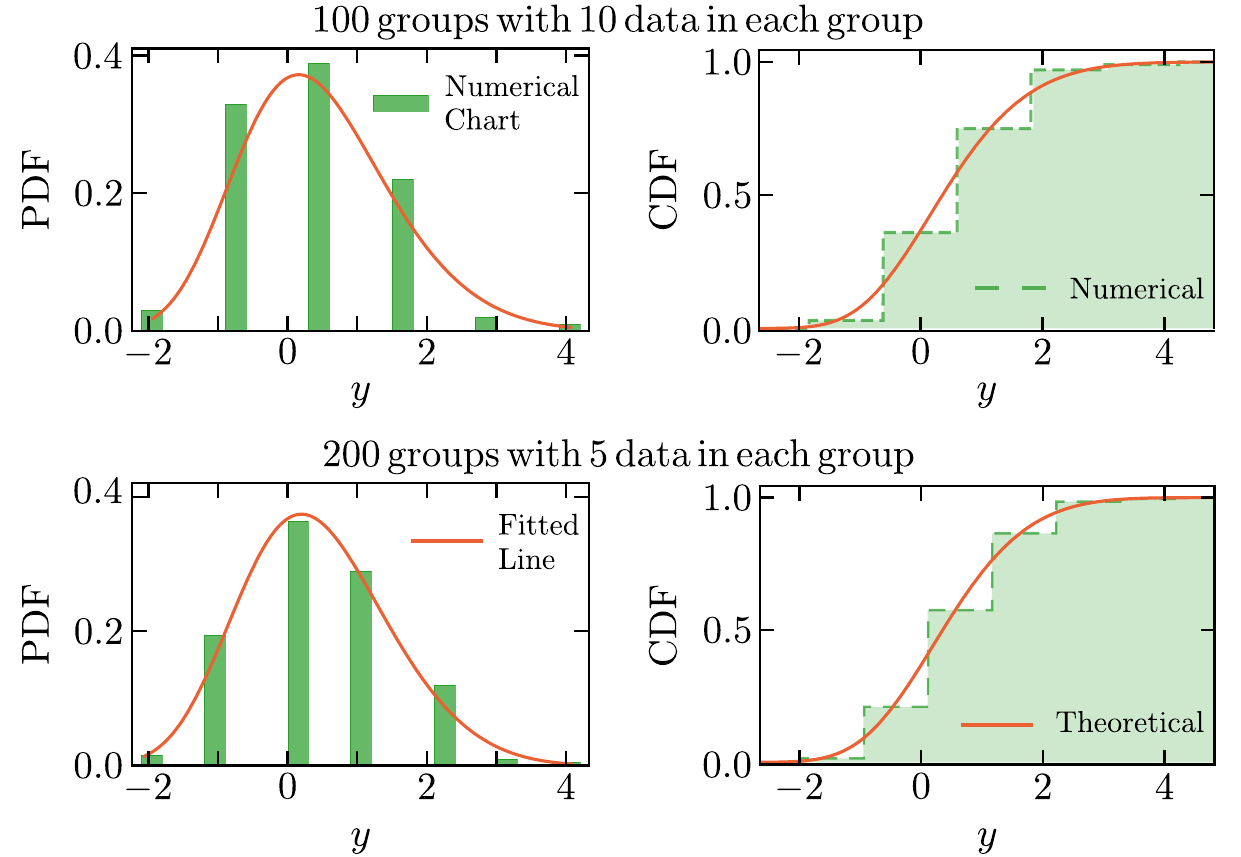}
\caption{Vortex-number GEV distribution for $\tau_Q=1000$. The numerical results  follow the Weibull distribution with an upper bound.}\label{tq1000}
\end{figure}

For further characterization,  we resort to large deviation theory and characterize the distribution of the maxima in long sequences of realizations.
According to the Fisher-Tippett-Gnedenko theorem\cite{Haan06}, the extreme values of the iid variables satisfy the generalized extreme value (GEV) distribution,
\be
G(x;\mu,\sigma,\xi)=
\begin{cases}
\exp\left(-\left(1+\frac{x-\mu}{\sigma}\xi\right)^{-1/\xi}\right), &\xi\neq0 \vspace{2mm}\\ 
\exp\left(-\exp\left(-\frac{x-\mu}{\sigma}\right)\right),&\xi=0.
\end{cases} 
\ee
with location parameter $\mu$, scale parameter $\sigma$ and shape parameter $\xi$.
The GEV distribution includes as limiting cases of the Weibull ($\xi<0$), Gumbel ($\xi\rightarrow 0$) and Fr\'echet ($\xi>0$) laws.
To determine the relevant GEV distribution, we use the block maxima method. Data from the stochastic trajectories is partitioned in blocks. In each block, the maximum vortex number $n_{M}$ is found and from the ensemble of blocks the probability ${\rm Prob}(n_{M}<x)$ is determined. 
We analyze the statistics of large vortex number deviations for slow quenches in the universal scaling regime in  Fig. \ref{tq1000}, that show the probability density function (PDF) and the cumulative distribution function (CDF) for their GEV distributions in different groups. Both PDF and CDF are shown as a function of the variable $y=\frac{x-\mu}{\sigma}$.
Specifically, for $\tau_Q=1000$,  data is partitioned into $100$ groups (top row in Fig. \ref{tq1000}) and $200$ groups (bottom row in Fig. \ref{tq1000}). The corresponding parameters are $\mu=13.056, \sigma=1.666, \xi=-0.149$ and $\mu=11.892, \sigma=1.943, \xi=-0.184$, respectively. Thus, GEV is  described by  the Weibull distribution, which has reflecting the existence of an upper bound to the vortex number. 
As shown in the Appendix~\ref{ExtreVal}, the validity of the Weibull distribution further extends away from the universal scaling regime and characterizes deviations also in the saturation regime associated with fast quench times.
\section{Discussion}
The characterization of the full distribution of topological defects generated across a phase transition is expected to have wide applications, ranging from condensed matter to quantum simulation and computation, and cosmology. Experimental efforts to date have focused on the universal power-law dependence of the mean number of defects with the quench time, which is successfully predicted by KZM.
Our results show that fluctuations away from the mean exhibit universality but are no longer captured by the KZM power-law scaling. The full counting statistics of vortices in a strongly coupled superconductor follows a universal binomial distribution. 
The distribution of maxima in a sequence of realizations is captured for the Weibull law in large deviation theory. The dependence of the tails of the distribution with the quench time dictates the suppression of topological effects, near or far from the onset of adiabatic dynamics. This dependence is thus crucial to analyze rare events associated with profusion or absence of topological defects. The complete suppression of topological defects is sought after in the preparation of novel phases of matter in quantum simulators and finite-time quantum annealing and quantum optimization.  It is also of relevance in a cosmological setting, as the observation of cosmic strings predicted by the KZM remains elusive.

\appendix
\section{Equations of motion for a holographic superconductor}\label{EQMov}
In the probe limit, the equations of motions for $\Psi$ and $A_\mu$ read,

\begin{eqnarray}\label{eomofwhole}
D_\mu D^\mu\Psi-m^2\Psi=0,~~~\nabla_\mu F^{\mu\nu}=-2\Im(\Psi^* D^\nu\Psi). 
\end{eqnarray}
in which $\Im(.)$ represents imaginary part. In explicit form,  these equations read 
\begin{eqnarray}\label{eomofwhole}
D_\mu D^\mu\Psi-m^2\Psi=0,~~~\nabla_\mu F^{\mu\nu}=-2\Im(\Psi^* D^\nu\Psi). 
\end{eqnarray}
in which $\Im(.)$ represents imaginary part. In explicit form,  these equations read 
\begin{eqnarray}
\label{eompsi}
\partial_t \partial_z \Phi - i A_t \partial_z \Phi - \frac12 [ i \partial_z A_t \Phi +\partial_z(f\partial_z\Phi)  - z \Phi 
+ (\partial_x^2 \Phi + \partial_y^2 \Phi) - i (\partial_x A_x + \partial_y A_y) \Phi &
\nonumber\\ - (A_x^2 + A_y^2) \Phi - 2 i (A_x \partial_x \Phi + A_y \partial_y \Phi) ] = 0,&
\\
\label{eom2}
\partial_t \partial_z A_t - (\partial_x^2 A_t + \partial_y^2 A_t) - f \partial_z (\partial_x A_x + \partial_y A_y) + \partial_t (\partial_x A_x + \partial_y A_y) 
+ 2 A_t |\Phi|^2 ~~~~~~&
 \nonumber\\+2f\Im(\Phi^* \partial_z \Phi)-2\Im(\Phi^* \partial_t \Phi)   = 0,&
\\
\label{eom3}
\partial_t \partial_z A_x - \frac12 \left[ \partial_z (\partial_x A_t + f \partial_z A_x) + \partial_y (\partial_y A_x - \partial_x A_y)+2\Im(\Phi^* \partial_x \Phi )  - 2 A_x |\Phi|^2 \right] = 0,&
\\
\label{eom4}
\partial_t \partial_z A_y - \frac12 \left[ \partial_z (\partial_y A_t + f \partial_z A_y) + \partial_x (\partial_x A_y - \partial_y A_x) +2\Im(\Phi^* \partial_y \Phi) - 2 A_y |\Phi|^2 \right] = 0,&
\\
\label{eom1}
\partial_z (\partial_x A_x + \partial_y A_y - \partial_z A_t)-2\Im(\Phi^* \partial_z \Phi)  = 0,&
\end{eqnarray}

where $\Phi=\Psi/z$. The above five equations are not independent, and their L.H.S. satisfy the following constraint equation
\begin{eqnarray}
-\frac{d}{dt}\text{Eq.\eqref{eom1}}-\frac{d}{dz}\text{Eq.\eqref{eom2}}+2\frac{d}{dx}\text{Eq.\eqref{eom3}}+2\frac{d}{dy}\text{Eq.\eqref{eom4}}\equiv4\Im(\text{Eq.\eqref{eompsi}}\times\Phi^*). 
\end{eqnarray}
 Therefore, there are four independent equations for four fields,  $\Phi, A_t, A_x$ and $A_y$. This also implies that our choice of the gauge $A_z=0$ is viable for the setup of the system. 
\section{Properties of binomial and Poissonian distributions restricted to even outcomes}\label{BinoEven}
\begin{figure}[t!]
\centering
\includegraphics[width=0.5\linewidth]{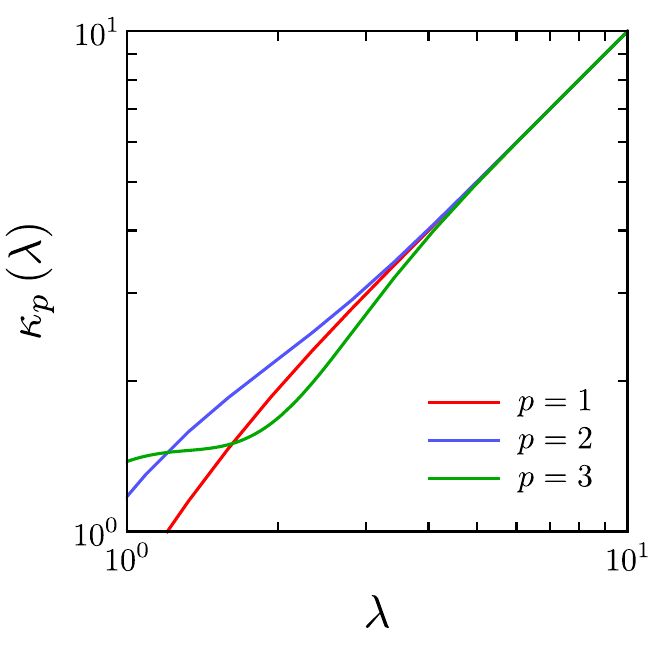}
\caption{Cumulants of the EP distribution as a function of the parameter $\lambda=Np$. For large values of $\lambda$, all cumulants approach the mean of the distribution $\langle n\rangle=\lambda\tanh (\lambda )$.}\label{figCumuEP}
\end{figure}

The even-binomial (EB) distribution is obtained by restricting to even outcomes the binomial distribution and is 
\be \label{evenbino}
P_{\rm EB}(n)=\frac{1}{A}\binom{N}{n}p^n(1-p)^{N-n},
\ee
where $N$ is the number of domains with broken symmetry, $p$ is the success probability to form a vortex, $n$ represents a given number of vortex and belongs to non-negative even integers, and $A=\frac{1+(1-2p)^N}{2}$ is the normalization constant. The first three cumulants of EB distribution are
\be
\kappa_1&=&Np\ \frac{1-(1-2p)^{N-1}}{1+(1-2p)^N},\label{k1}\\
\kappa_2&=&\frac{Np(1-p)}{\left(1+(1-2p)^N\right)^2} \bigg[1-(1-2p)^{2N-2}+4(N-1)(p-p^2)(1-2p)^{N-2}\bigg],\label{k2}\\
\kappa_3&=&\frac{N p(1-p)}{\left(1+(1-2 p)^N\right)^3}
 \bigg[1-2p-(1-2 p)^{3
   N-3}\nonumber\\
   &&+\left(1-4 (1-p) p (1-(N-1) (3-2 (N+4)
   (1-p) p))\right) (1-2 p)^{N-3}\nonumber\\
   &&
   -(1-4 (1-p) p (1+(N-1)
   (3+2 (N-2) (1-p) p))) (1-2 p)^{2 N-3}\bigg].\label{k3}
\ee
These cumulants satisfy a recursion relation such as $\kappa_{q+1}=p(1-p)\frac{d\kappa_q}{dp}$. In the limit of $N\to\infty$ and keeping the parameter $Np=\lambda$ finite, we get the even-Poisson (EP) distribution 
\be\label{peven}\lim_{\substack{N\to\infty \\ Np=\lambda}}
P_{\rm EB}(n)=\frac{2 e^{\lambda } \lambda ^n}{\left(e^{2 \lambda
   }+1\right) \Gamma (n+1)}={\rm sech}(\lambda)\frac{\lambda^n}{n!}.
\ee
In addition, in this limit the first three cumulants Eqs.\eqref{k1}, \eqref{k2} and \eqref{k3} become
\be
&&\lim_{\substack{N\to\infty \\ Np=\lambda}}\kappa_1 = \lambda \tanh(\lambda),\label{kk1}\\
&&\lim_{\substack{N\to\infty \\ Np=\lambda}}\kappa_2 = \lambda  \left[\tanh (\lambda )+\lambda 
   \text{sech}^2(\lambda )\right],\label{kk2}\\
 &&\lim_{\substack{N\to\infty \\ Np=\lambda}}\kappa_3 = \lambda  \left[\tanh (\lambda )+\lambda  (3-2 \lambda
    \tanh (\lambda )) \text{sech}^2(\lambda )\right]. \label{kk3}
\ee
In the large $\lambda$ limit, these three cumulants Eqs. \eqref{kk1}, \eqref{kk2} and \eqref{kk3} are close to each other, i.e., $\kappa_1\approx\kappa_2\approx\kappa_3\approx \lambda$. From the explicit expressions for the cumulants and the behavior of the $\tanh(\lambda)$ function one can see that for values of $\lambda\geq 4 $ this is already satisfied to great accuracy, see Fig.  \ref{figCumuEP}.
Cumulants of the EP distribution for large parameter $\lambda$ approach those of the (unrestricted) Poissonian distribution with mean $\lambda\tanh (\lambda )\approx \lambda$. The regime of quench rates explored in the main extends to lower values of $\lambda$ in which deviations of the third cumulant from the asymptotic form occur (for any number of trajectories).

\section{Full counting statistics of vortices: numerical simulations for a holographic superconductor and even-Poissonian distribution}\label{pk}
\begin{figure}[h!]
\centering
\includegraphics[width=1\linewidth]{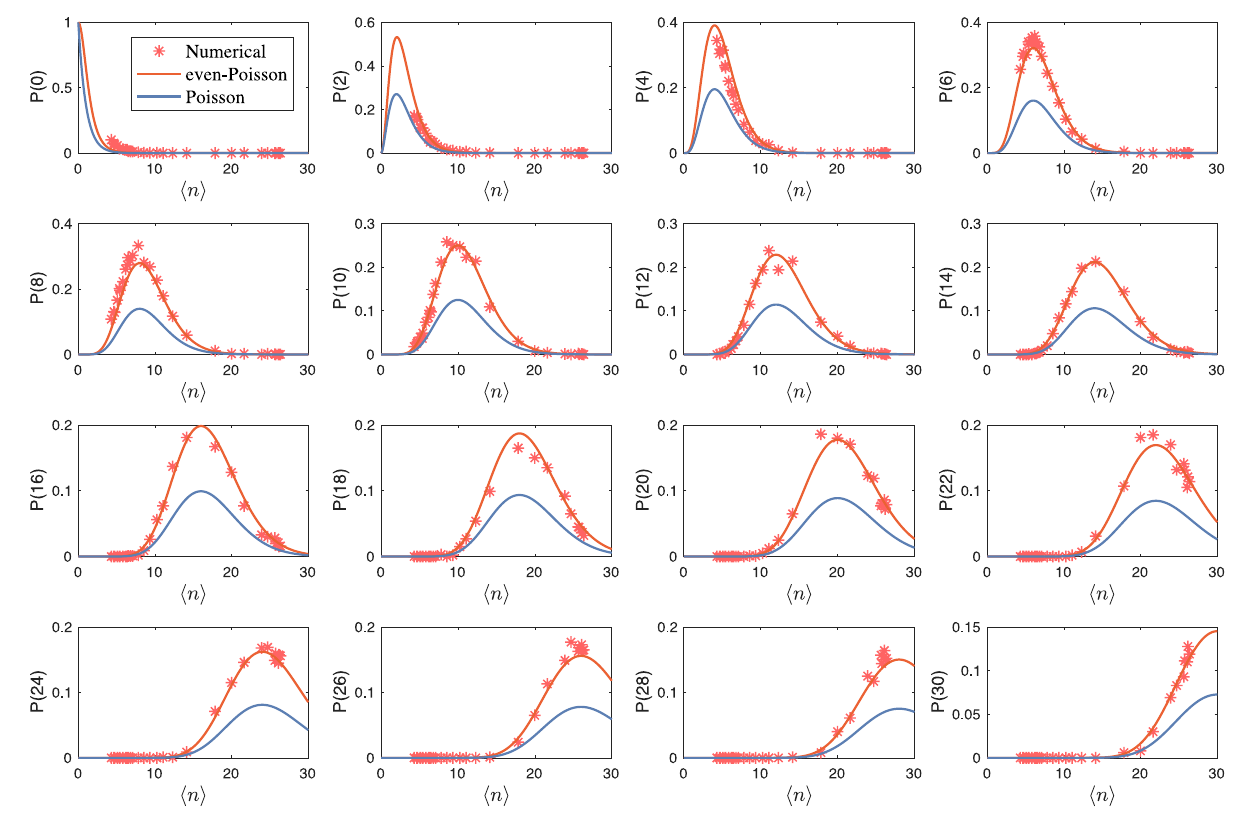}
\caption{Probability distribution of each even appearance of vortices $P(n=0,2,4,\dots, 30)$ with respect to the average vortex number $\langle n\rangle$. Except for $n=0$, other distributions $P(n)$ satisfy the EP distribution very well. 
}\label{pndirectjoin}
\end{figure}

In Fig.  \ref{pndirectjoin}, we compare the numerical results of the probability distribution $P(n=0,2,4,\dots, 30)$ of a given vortex number $n$ with respect to theoretical value of the EP distribution,  and the  (unrestricted) Poisson distribution, which is a limit of (unrestricted) Binomial distribution with $N\to\infty$ and keeping the parameter $Np=\lambda$ finite. The agreement between the numerical simulations and the EP distribution is excellent provided sufficient statistics.

\begin{figure}[h!]
\centering
\includegraphics[width=0.5\linewidth]{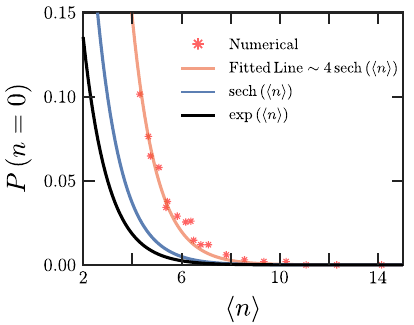}
\caption{Probability distribution of $P(n=0)$ with respect to the average number of vortices $\langle n\rangle$. Numerical results are consistent with theoretical EP distribution (red line) up to an $\mathcal{O}(1)$ factor. Black line is the (unrestricted) Poissonian distribution.}\label{P0N}
\end{figure}

The probability $P(n=0)$ to observe no vortices at all is a rare event away from the adiabatic limit and can be estimated from the 
 EP distribution. Using Eq.   \eqref{peven}, it reads 
\be\label{p0N}
P_{\rm EP}(n=0)={\rm sech}(\lambda).
\ee
Numerical results for $P(n=0)$ with respect to $\langle n\rangle$ are shown in Fig.  \ref{P0N} with $P(n=0)\approx3.998\times\text{sech}(\langle n\rangle)$.  The value of the prefactor is found to decrease with an increasing number of trajectories of simulations. This implies that the error between the numerical results and the theoretical EP distribution is due to the limited sampling statistics in the simulations. By increasing it, the numerical results are expected to approach the theoretical EP distribution.  

\section{Large fluctuations and cumulative probability in the tails of the vortex-number distribution as function of the  quench time}

\begin{figure}[h!]
\centering
\includegraphics[width=1\linewidth]{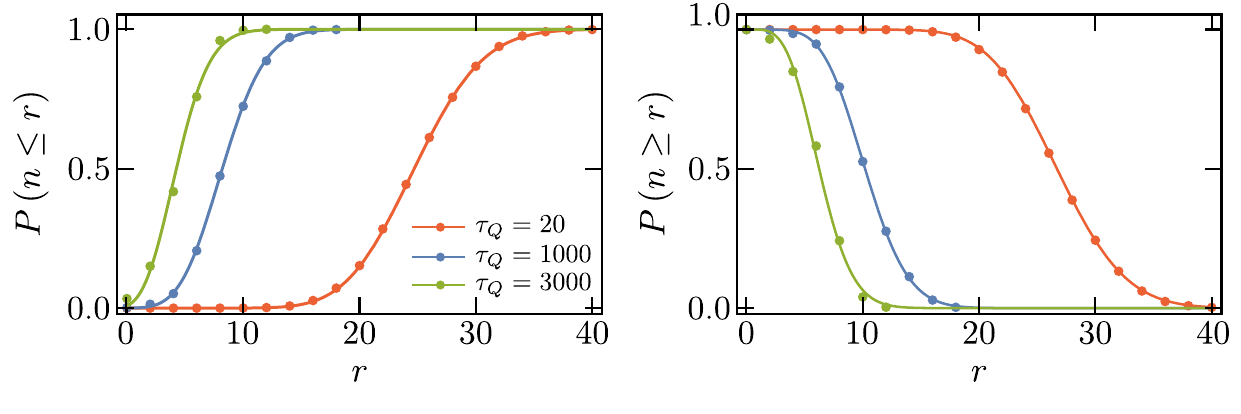}
\caption{Cumulative probability  $P(n\leq r)$ (Left) and $P(n\geq r)$ (Right)  for $\tau_Q=20, 1000$ and $3000$. Numerical results fit closely the theoretical predictions derived from the EP distribution in Eq. \eqref{pebksr} and Eq. \eqref{pebkgr}. }\label{cumupsmall}
\end{figure}

In the main text, we show the cumulative probability of even-Poissonian distribution for $P_{\rm EP}(n\leq r)$ and $P_{\rm EP}(n\geq r)$ as
\be\label{pebksr}
P_{\rm EP}(n\leq r)&=&1-\frac{\text{sech}(\lambda ) \lambda ^{2
   \left\lfloor \frac{r}{2}\right\rfloor +2} \,
   _1F_2\left(1;\left\lfloor \frac{r}{2}\right\rfloor
   +\frac{3}{2},\left\lfloor \frac{r}{2}\right\rfloor
   +2;\frac{\lambda ^2}{4}\right)}{\left(2
   \left(\left\lfloor \frac{r}{2}\right\rfloor
   +1\right)\right)!}, \\
P_{\rm EP}(n\geq r)&=& \frac{\text{sech}(\lambda ) \lambda ^r \,
   _1F_2\left(1;\frac{r}{2}+\frac{1}{2},\frac{r}{2}+1
   ;\frac{\lambda ^2}{4}\right)}{r!}. \label{pebkgr}
\ee
where $\, _1F_2$ is a hypergeometric function
and $\lfloor\cdot\rfloor$ is the floor function. In Figure \ref{cumupsmall},  the numerical results of the cumulative probabilities are shown to match very well the theoretical predictions for a broad range of rates at which the transition is crossed, ranging from slow quenches with  $\tau_Q=3000$ to the fast quench limit with $\tau_Q=20$.

\subsection{Chernoff bound}

\begin{figure}[h!]
\centering
\includegraphics[width=1\linewidth]{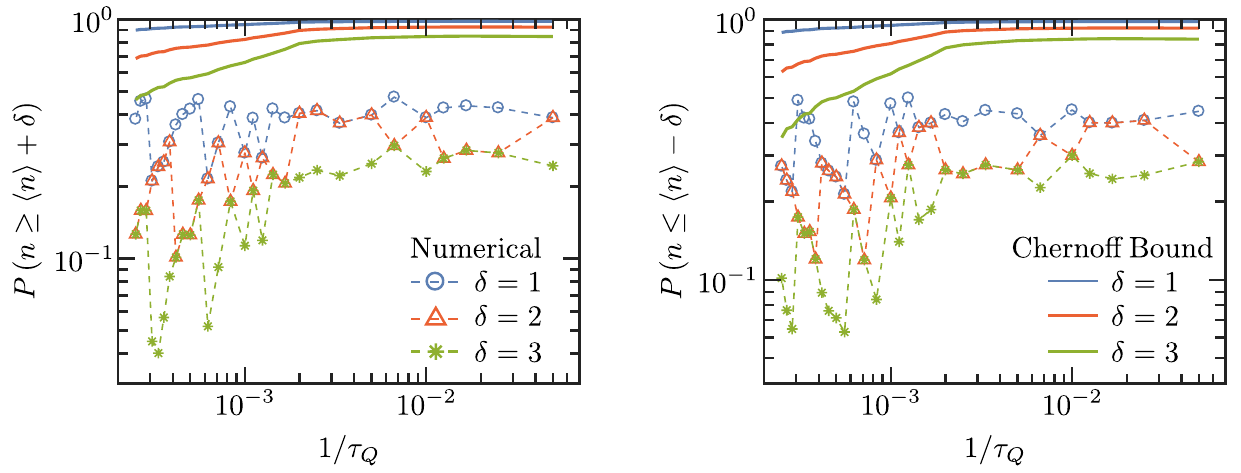}
\caption{The bound of the upper tail (Left) and lower tail (Right) of distributions of vortex numbers as a function of the inverse of quench time. Numerical results satisfy the theoretical bounds, that capture the dependence on the quench time. }\label{bound}
\end{figure}

The Chernoff bound \cite{Molloy} can be used to derive exponentially decreasing bounds on the tail distributions of vortex numbers.  In its looser form, the Chernoff bound can be written as 
\be
&&P(n\leq\langle n\rangle-\delta)\leq e^{-\frac{\delta^2}{2\langle n\rangle}}, ~~~~~\text{(Lower tail)}\\
&&P(n\geq\langle n\rangle+\delta)\leq e^{-\frac{\delta^2}{2\langle n\rangle +\delta}}. ~~~~\text{(Upper tail)}
\ee
In Fig. \ref{bound} we plot bound of lower tail and upper tail for $\delta=1,2,3$. The numerical results satisfy the Chernoff bound very well. The latter can thus be used to capture the dependence on the quench time of the large fluctuations away from the mean, associated with the tails of the distribution.

\section{Extreme value distribution of vortex numbers}\label{ExtreVal}
According to the Fisher-Tippett-Gnedenko theorem \cite{Haan06}, the extreme maximal values of the {\it independently and identically distributed} (iid) variables satisfy the {\it generalized extreme value} (GEV) distribution, 
\be
G(x;\mu,\sigma,\xi)=
\begin{cases}
\exp\left(-\left(1+\frac{x-\mu}{\sigma}\xi\right)^{-1/\xi}\right), &\xi\neq0 \vspace{2mm}\\ 
\exp\left(-\exp\left(-\frac{x-\mu}{\sigma}\right)\right),&\xi=0.
\end{cases} 
\ee
in which, $\mu$ is the location parameter, $\sigma$ is the scale parameter and $\xi$ is the shape parameter. Note that $\xi (x-\mu)/\sigma+1>0$ and zero otherwise. The above GEV distribution function $G(x;\mu,\sigma,\xi)$ is the cumulative density function (CDF), whose corresponding probability density function (PDF) can be written as 
\be
P(x;\mu,\sigma,\xi)=\begin{cases}
 \frac{1}{\sigma}{\left(\frac{\xi  (x-\mu
   )}{\sigma }+1\right)^{-\frac{1}{\xi }-1}\exp\left({-\left(\frac{\xi  (x-\mu )}{\sigma
   }+1\right)^{-1/\xi }}\right) }, &
   \xi \neq 0  \vspace{2mm}\\
   \frac{1}{\sigma }{\exp\left({-\frac{x-\mu }{\sigma }-\exp\left({-\frac{x-\mu
   }{\sigma }}\right)}\right)}, & \xi =0.
\end{cases}
\ee
If $\xi<0$, the GEV distribution is called Weibull distribution which is upper bounded. if $\xi=0$, the GEV distribution is called Gumbel distribution which has a light tail. Finally, if $\xi>0$, the GEV distribution is called Fr\'echet distribution which has a heavy tail and a lower bound.\\
\\  
\begin{figure}[h!]
\centering
\includegraphics[width=0.9\linewidth]{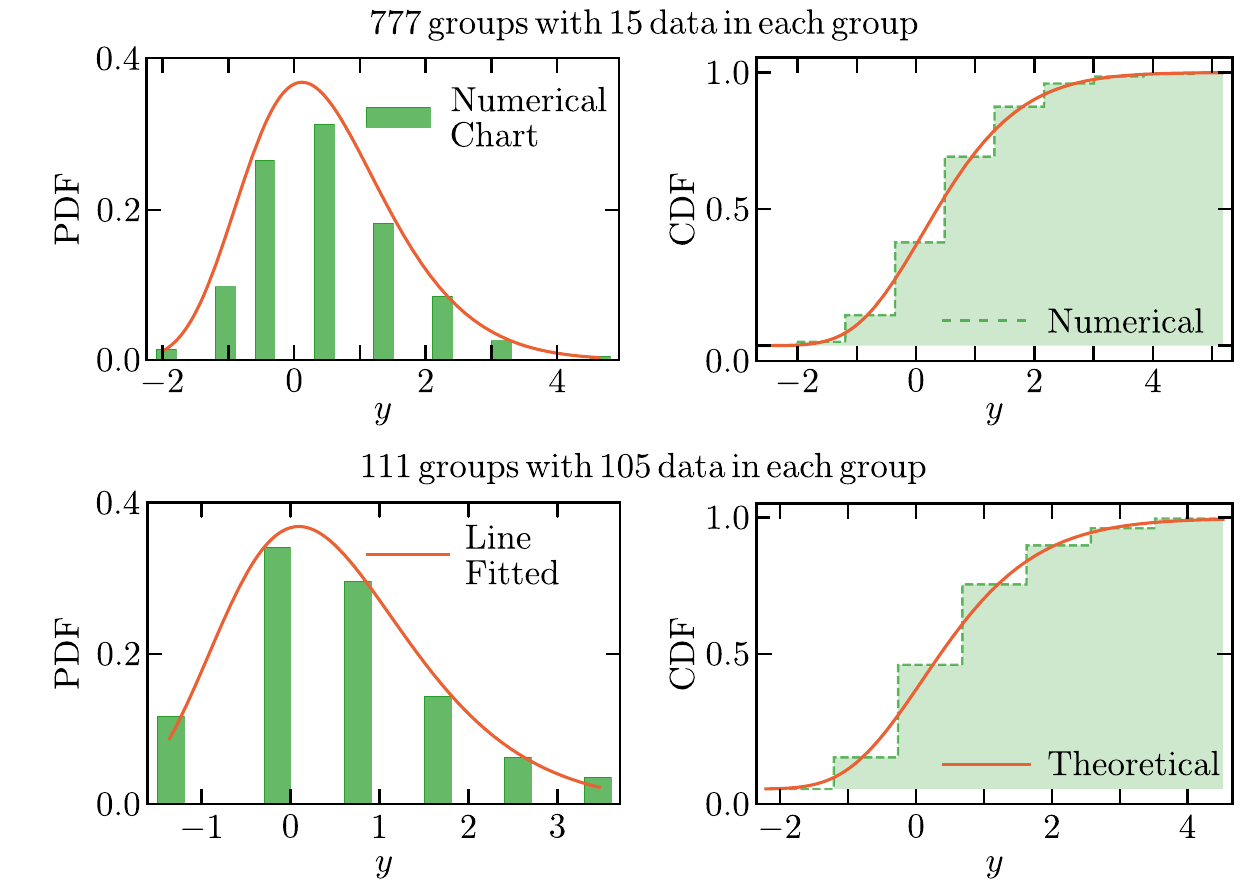}
\caption{Probability density function (PDF) and cumulative distribution function (CDF) of extreme maximum values of the vortex number spontaneously generated in the limit of fast quenches with  $\tau_Q=20$. Independently of the partitioning size, he data is well described by a Weibull distribution with an upper bound.  }\label{tq20}
\end{figure}

In practice, to analyze the extreme value distributions for iid variables, it is customary to separate the data into several groups (or blocks), and then proceed to identify the maximum in each group. The final list of maxima will tend to satisfy the above GEV distribution. This method is called `Block Maxima' method, and we adopt it to study the maximum value distributions for the vortex numbers in numerical simulations. There are some arbitrary choices in the partition of the data. We partition the data into more than $100$ groups, which is sufficient for the observed vortex-number maxima distribution to be identified with the GEV. \\
\\
The extreme maximal value distributions of the vortex number for a slow quench (such as $\tau_Q=1000$) are shown in the main text. Here, we show the PDF and CDF of the maximum values for a fast quench ($\tau_Q=20$) in Fig. \ref{tq20}. Specifically, for $\tau_Q=20$ we have $11655$ numerical data of the vortex number.  They are partitioned into $777$ groups (top row in Fig. \ref{tq20}) and $111$ groups (bottom row in Fig. \ref{tq20}). Both PDF and CDF are shown as a function of the variable $y=\frac{x-\mu}{\sigma}$. In the top row of Fig. \ref{tq20}, the parameters are $\mu=32.841, \sigma=2.382, \xi=-0.113$; while in the bottom row the parameters are $\mu=36.558, \sigma=2.114, \xi=-0.090$. The analysis of the data based on these partitions shows that the GEV distribution of the maxima for $\tau_Q=20$ belongs to Weibull distribution, which means there is an upper bound. 
\newpage
\acknowledgments
We acknowledge funding support from the Spanish MICINN (PID2019-109007GA-I00) and the National Natural Science Foundation of China (Grants No. 11675140, 11705005 and 11875095).

\bibliographystyle{JHEP}
\bibliography{Vortex_FCS}

\end{document}